\documentclass{article}

\usepackage{arxiv}

\usepackage[utf8]{inputenc} 
\usepackage[T1]{fontenc}    
\usepackage{hyperref}       
\usepackage{url}            
\usepackage{booktabs}       
\usepackage{amsfonts}       
\usepackage{nicefrac}       
\usepackage{microtype}      
\usepackage{lipsum}		
\usepackage{amssymb,amsmath}
\usepackage{listings}
\usepackage{graphicx}
\usepackage{subfig}
\usepackage{verbatim}

\title{Decentralised, privacy-preserving Bayesian inference for mobile phone contact tracing}

\author{
  Daniel Tang\\
  Leeds Institute for Data Analytics\thanks{This project has received funding from the European Research Council (ERC) under the European Union’s Horizon 2020 research and innovation programme (grant agreement No. 757455)}\\
  University of Leeds\\
  Leeds, UK\\
  \texttt{D.Tang@leeds.ac.uk} \\
}

\begin{document}
\maketitle

\begin{abstract}
Many countries are currently gearing up to use smart-phone apps to perform contact tracing as part of the effort to manage the COVID-19 pandemic and prevent resurgences of the disease after the initial outbreak. With the announcement of the Apple/Google partnership to introduce contact-tracing functionality to iOS and Android\cite{applegoogle}, it seems likely that this will be adopted in many countries.

An important part of the functionality of the app will be to decide whether a person should be advised to self-isolate, be tested or end isolation. However, the privacy preserving nature of the Apple/Google contact tracing algorithm means that centralised curation of these decisions is not possible so each phone must use its own ``risk model'' to inform decisions. Ideally, the risk model should use Bayesian inference to decide the best course of action given the test results of the user and those of other users. Here we present a decentralised algorithm that estimates the Bayesian posterior probability of viral transmission events and evaluates when a user should be notified, tested or released from isolation while preserving user privacy. The algorithm also allows the disease models on the phones to learn from everyone's contact-tracing data and will allow Epidemiologists to better understand the dynamics of the disease.

The algorithm is a message passing algorithm, based on belief propagation, so each smart-phone can be used to execute a small part of the algorithm without releasing any sensitive information. In this way, the network of all participating smart-phones forms a distributed computation device that performs Bayesian inference, informs each user when they should start/end isolation or be tested and learns about the disease from user's data.
\end{abstract}


\section{Introduction: the need for Bayesian inference}

While much has been said about how to use smart-phones to inform people that they have been in close proximity to an infected person\cite{dp3t}\cite{applegoogle}\cite{pepppt}, relatively little has been said about how to evaluate the best course of action given information about one's own test results and those of one's contacts.

Since clinical tests for SARS-CoV-2 have been found to have quite high rates of false-negatives\cite{fang2020sensitivity}, accurately diagnosing whether someone is infected isn't as easy as just testing them. However, by taking into account symptoms and test results of a user \textbf{and} of the user's close contacts we can potentially increase the sensitivity of the diagnosis. For example, suppose a contact of a confirmed case is symptomatic but tests negative, yet on the next day a contact of his independently becomes symptomatic and tests positive. If the newly confirmed case was in contact only with the negative case around the inferred time of infection, we may have enough information to deem the negative test a false-negative and contact trace/isolate them in any case.

Also, there is evidence that there is a high proportion of asymptomatic transmission of SARS-CoV-2\cite{lavezzo2020suppression}. This means that it is important to identify asymptomatic carriers. So, if someone is confirmed infected, it is important to trace who they caught the disease from, as well as who they may have passed the disease on to, so that a potential asymptomatic carrier can be identified and their contacts traced. For example, if two people become symptomatic and test positive, we know that each of them must have been infected a few days before their respective symptoms began. Each person is likely to have had a number of contacts in that window of time but if they have one contact in common then it becomes much more probable that the common contact was the original source of the infection. If it turns out the suspected source was symptomatic around 5 days ago (but opted not to be tested) then this evidence may be enough to immediately trace their contacts to identify other possible infectees (it would be less important to isolate the original source at this point as they may be less infectious at this point. However, if that person had a tendency to have a very large number of contacts a day and the timing meant there was still a chance of residual infectiousness, it may be deemed prudent to isolate for a number of days).

Finally, if a person is traced as a contact of a confirmed case, but the contact occurred a few hours ago, there is little point in testing that person immediately as the test is unlikely to be positive even if the person will go on to develop the disease. There is also no need for the person to immediately isolate as they are probably not infectious at this point. However, at some time in the near future we will want to test and possibly isolate that person, but when is the optimal time to do this?

Doing this kind of inference automatically, while maintaining the privacy of a user's close contacts, requires a reasonably sophisticated inference algorithm. Simply alerting close contacts of confirmed cases is not enough. Here we present an algorithm that uses a modification of the belief propagation algorithm\cite{pearl2014probabilistic} to do Bayesian inference in order to decide on a testing schedule and isolation timings that minimise expected overall cost. The algorithm also allows the disease model to learn from user's data and allows Epidemiologists to gain information about the dynamics of the disease.

\section{Description of the problem}

Suppose mobile-phone contact tracing is in operation using the Apple/Google contact tracing specification\cite{applegoogle}. Each day, enabled phones randomly choose a ``daily exposure key''. When enabled phones come close they exchange proximity IDs (PIDs) which are numbers derived from their daily exposure key. The phones keep a log of the exchanged PIDs along with the time of the contact and their own daily exposure keys. If a person is deemed at risk, a subset of their daily exposure keys are published on a public server along with a payload containing some data. Each day, everyone's phone checks for newly published exposure keys and checks them against the PIDs of their close contacts. If there is a match, the phone learns the user has been in close proximity to a person at risk.

The problem we consider here is how to calculate whether and when to notify the user to self-isolate, perform tests and release from isolation, while maintaining the privacy of a user's close contacts. In addition, we would like to use user's contact and test result data to update a set of epidemiological parameters about the natural history of the disease, again without compromising the privacy of the users contacts.

\section{Expressing the joint probability over transmission events}

If two people, A and B, come within close proximity and broadcast PIDs $a$ and $b$ respectively, then there is a chance $P(\tau_{a \rightarrow b})$ that A will transmit the virus to B and a chance $P(\tau_{b \rightarrow a})$ that B will transmit the virus to A. Our first aim is to calculate the joint probability of all transmission events and all the tests/observations that have been done on all people to date. When we talk about transmission, we mean the transfer of viruses from one person to another such that if the other person was susceptible, they would become infected. Transmission to an already infected person is still a transmission event, but has no effect.

The binary random variables $\tau_{a\rightarrow b}$ take values either 1 or 0. A value of 1 denotes the event that a transmission occurred from A to B, while a value of 0 denotes the event that A and B had a close contact but no transmission occurred from A to B.

When a contact first occurs, the user has no information about the other participant so the probability of transmission from others is set to a prior probability. Up-to-date regional prior probabilities can be downloaded from publicly available servers and the phone can estimate its current region from recent events such as the name/ID of the most recently connected cell tower (this information will always stay local on the phone).

\subsection{The disease model}

Consider the phone of person A at time $t$ with a log of contacts $C(t) = \left< t_{1},a_1,b_1 \right> ... \left< t_{n},a_n,b_n \right>$, where $t_i$ is the time of the contact relative to the start of the log and $a_i$ and $b_i$ are the exchanged PIDs.

Define the \textit{first-exposure time} to be the earliest time that A was exposed to the virus and let $\epsilon(t)$ be the event that the first exposure is at time $t$. We assume that there is a continuous and slowly changing risk of exposure from the environment (e.g. from surfaces or close contacts that were missed for any reason) at a rate $\rho$. $\rho$ can be calculated from the local prevalence of the disease, the uptake of the app and the rate of contacts of the user. Given this, the probability density of first exposure since the start of the log is
\begin{equation}
P\left(\epsilon(t)|C(t)\right) = \left(\rho + \sum_{i=1}^n \delta(t - t_i)\tau_{b_i\rightarrow a_i} \prod_{j<i}(1-\tau_{b_j\rightarrow a_j})\right) e^{-\rho t}
\label{firstExposure}
\end{equation}
where $\delta$ is the Dirac delta function.

Define the \textit{disease onset time} to be the earliest time that an infected person has a non-zero probability of transmitting the disease and let $\omega(t)$ be the event that disease onset occurs at time $t$. If we assume that there was no infection at the start of the contact log then
\begin{equation}
P(\omega(t)|C(t)) = \int_0^{t} P(\omega(t)|\epsilon(t_\epsilon))P(\epsilon(t_\epsilon)|C(t)) dt_\epsilon
\end{equation}
where $P(\omega(t)|\epsilon(t_\epsilon))$ represents the incubation time of the disease. This could be an analytical distribution such as a Weibull distribution $P_w(t-t_\epsilon)$, or some more complex model.

After disease onset, the amount of viral shedding will vary with time. Let $\iota(t) = \rho_i$ be the event that a person is shedding viruses at a rate $\rho_i$ at time $t$, then the probability density of viral shedding is
\begin{equation}
P(\iota(t) = \rho_i|C(t),\alpha) = \int_0^t \delta(\iota(t|\omega(t_\omega),\alpha)-\rho_i)P(\omega(t_\omega)|C(t)) dt_\omega
\end{equation}
where $\iota(t|\omega(t_\omega),\alpha)$ represents the clinical course of the disease and $\alpha$ is a binary random variable indicating whether the person is an asymptomatic carrier (in which case, their infectiousness curve may be different from symptomatics). This could be a Beta distribution $P_\beta(t-t_\omega)$ or something more complex. Asymptomatics may have a completely different distribution, or may have the same distribution multiplied by a constant $\iota_\alpha$.

Given a person's infectiousness, we assume that a close contact is a Bernoulli draw, based on that person's infectiousness.
\begin{equation}
P(\tau|\iota(t)) = B(\tau, \iota(t))
\end{equation}
If detailed information about the contact is logged (e.g. duration and proximity), the Bernoulli distribution could be replaced with a more sophisticated model of the contact to get a more accurate probability of transmission (e.g. \cite{sattler2020risk}). Here we will assume a Bernoulli distribution to demonstrate the algorithm, but the same development could be applied to a more complex transmission model.

Suppose we have a number of tests $\xi_1...\xi_n$ that give us information about whether a person is infected with SARS-CoV-2. The characteristics of the $n^{th}$ test are defined by it's specificity, $P(\xi^-_n|\bar{i})$, which gives the probability that a person will test negative on test $n$ given that they are not infected, and it's sensitivity $P(\xi^+_n(t)|\omega(t_\omega),\alpha)$ which gives the probability that a person tests positive on test $n$ at time $t$ given that they had disease onset at time $t_\omega$ and whether they are asymptomatic carriers.

Tests may take the form of clinical tests (e.g. PCR tests, antigen tests or antibody tests) but may also be simple observations of the presence or absence of symptoms (e.g. fever, persistent-cough or loss of taste).

The conditional probability of a positive test result is given by
\[
P(\xi^+_n(t)|C(t),\alpha) = \int_0^t P(\xi^+_n(t)|\omega(t_\omega),\alpha)P(\omega(t_\omega)|C(t)) dt_\omega + \left(1-\int_0^tP(\omega(t_\omega)|C(t)) dt_\omega\right)(1-P(\xi^-_n|\bar{i}))
\]

\subsection{The joint probabilities}

The above conditional probabilities for each individual can be put together to create the joint probability over all people.

The probability that a contact with a person at time $t$ ends in a transmission event from that person is independent of other variables given the transmission events to that person before $t$, so we can imagine building up the joint from conditionals starting with the earliest contact and moving one at a time to the latest. If the $i^{th}$ user has a log of contacts $C_i$ and we let $\mathcal{C} = \cup_{i=1}^N C_i$ be the set of all close contacts in the logs of all people, and $\mathcal{T}$ be the set of all transmission events, it can be seen that the joint probability is given by
\begin{equation}
P(\mathcal{T},\theta) =
P(\theta)
\prod_{\left<t,a,b\right> \in \mathcal{C}}
P\left(\tau_{a\rightarrow b}|
  \left\{ \tau_{b' \rightarrow a} :
    \left<t',b',a\right> \in \mathcal{C} \wedge
    t' < t
  \right\},\theta
\right)
\label{allJoint}
\end{equation}
where $\theta$ is the set of epidemiological parameters that specify, for example, the incubation period distribution and the course of viral shedding, and $P(\theta)$ are our prior beliefs in those parameter values based on current scientific understanding.

We can group the terms in equation \ref{allJoint} by onward transmission from each user, giving

\[
P(\mathcal{T},\theta) =
P(\theta)\prod_{i}
\prod_{\left<t,a,b\right> \in C_i}
P\left(\tau_{a\rightarrow b}|
  \left\{ \tau_{b'\rightarrow a'}:
    \left<t',a',b'\right> \in C_i \wedge
    t' < t 
  \right\},\theta
\right)
\]

If we note that the probability of a positive/negative test result also depends only on the user's contacts prior to the test and whether the user is an asymptomatic carrier, then we can also add test results to the joint. If $T_i$ is the set of tests taken by user $i$,  and $\alpha_i$ indicates whether they are asymptomatic carriers, the joint becomes
\begin{equation}
P(\mathcal{T}, T_{1..N}, \alpha_{1..N},\theta) =
P(\theta)
\prod_{i} P(\alpha_i)
\prod_{\left<t,a,b\right> \in C_i}
P\left(\tau_{a\rightarrow b}|\mathcal{L}(C_i,t),\alpha_i, \theta \right)
\prod_{\xi_{n}(t) \in T_i}
P(\xi_{n}(t)|\mathcal{L}(C_i,t),\alpha_i)
\label{joint}
\end{equation}
where $P(\alpha)$ is the prior probability of being an asymptomatic carrier, and
\[
\mathcal{L}(C,t) =   \left\{ \tau_{b\rightarrow a}:\left<t',a,b\right> \in C \wedge t' < t \right\}
\]
This joint can be expanded to include each user's infectiousness at the time of each contact. For ease of notation, we imagine an ``augmented contact log'' which  includes the infectiousness in the Tuple of each contact so $\left<t,a,b,\iota\right> \in C_i$ is a contact at time $t$ where the broadcaster of PID $a$ had infectiousness $\iota$. So, the joint now looks like this
\begin{equation}
P(\mathcal{T},\mathcal{I}, T_{1..N}, \alpha_{1..N},\theta) =
P(\theta)
\prod_{i} P(\alpha_i)
\prod_{\left< t,a,b, \iota \right> \in C_i}
B(\tau_{a\rightarrow b}|\iota)
P\left(\iota|\mathcal{L}(C_i,t),\alpha_i,\theta \right)
\prod_{\xi_{n}(t) \in T_i}
P(\xi_{n}(t)|\mathcal{L}(C_i,t),\alpha_i)
\label{ijoint}
\end{equation}
where $\mathcal{I}$ is the set of all infectiousnesses.

\section{Calculating posterior marginal probabilities of transmission events}

\begin{figure}
\begin{center}
\includegraphics[width=10cm]{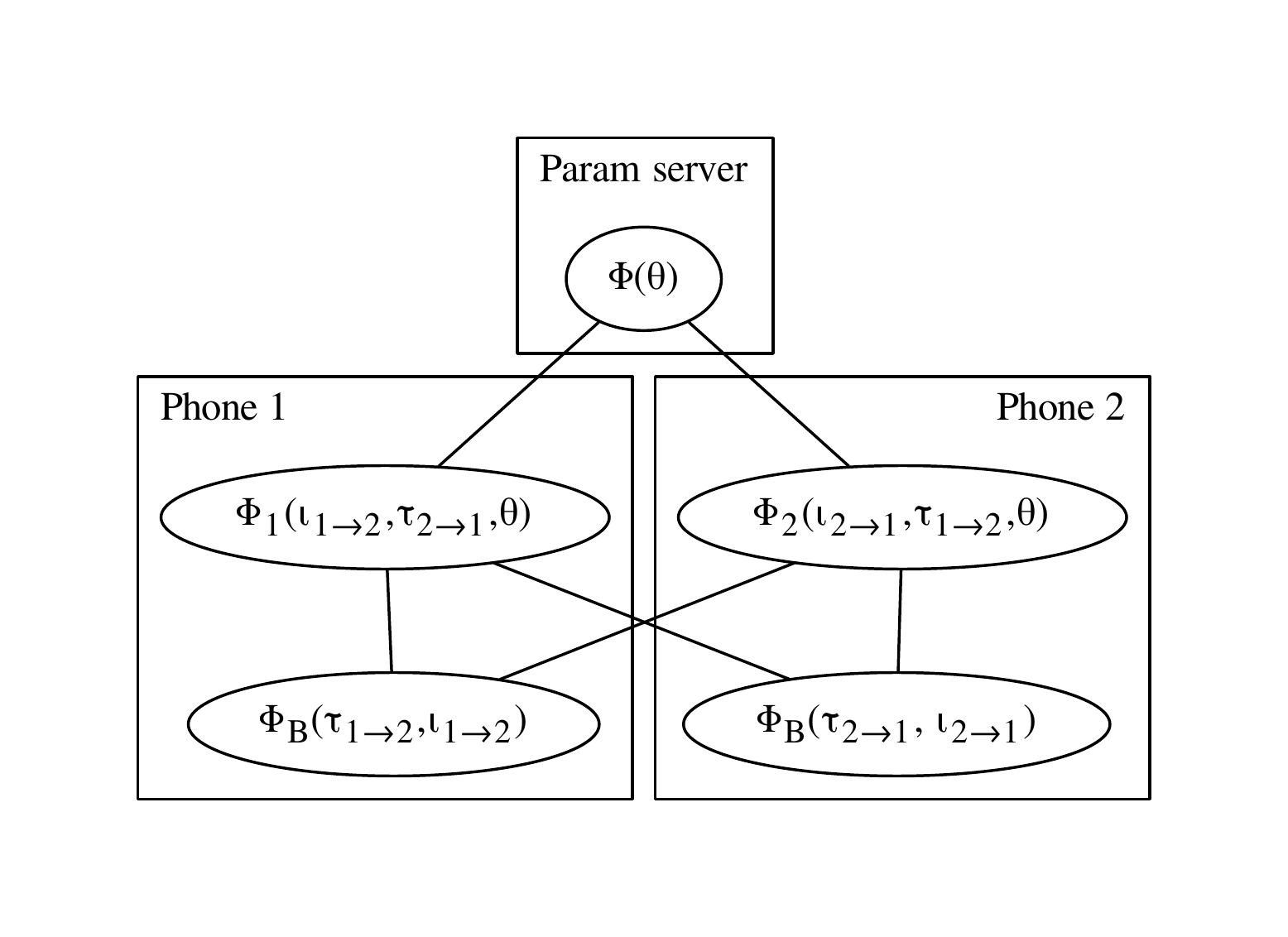}
\end{center}
\caption{Layout of cluster graph for two phones that have exchanged PIDs}
\label{ClusterGraph}
\end{figure}

We can now calculate the posterior marginal probabilities of all transmission events, given all the test results. We do this by embedding the joint in a cluster graph and performing belief propagation\cite{pearl2014probabilistic}\cite{koller2009probabilistic}. Figure \ref{ClusterGraph} shows the layout of the cluster graph for two phones that have exchanged PIDs. Each phone, $i$, has authority over a cluster that represents the disease model of the user
\begin{equation}
\Phi_i(\iota_{1..j},\tau_{1..j}, \theta) =
\sum_{\alpha_i}P(\alpha_i)
\prod_{\left< t,a,b, \iota_{a\rightarrow b} \right> \in C_i}
P\left(\iota_{a\rightarrow b}|\mathcal{L}(C_i,t),\alpha_i, \theta \right)
\prod_{\xi_{n}(t) \in T_i}
P(\xi_{n}(t)|\mathcal{L}(C_i,t),\alpha_i)
\label{ijoint}
\end{equation}
where $\iota_{1..j} = \left\{\iota_{a\rightarrow b}:\left< t,a,b, \iota \right> \in C_i\right\}$ and $\tau_{1..j} = \left\{\tau_{b\rightarrow a}:\left< t,a,b, \iota \right> \in C_i\right\}$.

Each phone also has authority over one Bernoulli cluster for each onward transmission event from the user
\begin{equation}
\left\{ \Phi_{B}(\tau_{a\rightarrow b},\iota_{a\rightarrow b}) = B(\tau_{a\rightarrow b}|\iota_{a\rightarrow b}) : \left<t,a,b,\iota\right> \in C_i \right\}
\end{equation}
 
Finally, there is a cluster containing the epidemiological model parameters which is held on a central server, from which all phones can obtain an up-to-date posterior distribution of the disease model parameters.

By splitting the joint into clusters in this way, we end up with a Bethe cluster graph which has good properties for performing belief propagation\cite{koller2009probabilistic}, so we would expect belief propagation to perform well on this graph both in terms of the accuracy of the results and the number of messages passed.

Given the cluster graph, belief propagation calculates the marginal probability of each cluster. The marginals can be conditioned on the results of all tests by simply setting the values of each test when calculating the messages to pass.

\subsection{Passing of messages}

We now describe how belief propagation can be achieved on the network of mobile phones, and the parameter server, without the need to expose any sensitive information.

Consider the messages sent from the $i^{th}$ phone. There are three types of message that need to be sent: ``forward'' messages from the phone's Bernoulli clusters to the $\Phi_j$ clusters of other phones, ``backward'' messages from the phone's $\Phi_i$ cluster to the Bernoulli clusters on other phones and ``epidemiological'' messages giving the parameter server information about the nature of the disease. We use the notation $\delta$ for forward messages and $\delta'$ for backward messages.

In order to reduce the total number of messages sent, messages are only sent if the information contained within them exceeds a threshold value. Since all edges that span phones represent binary variables, the information content of a message between phones (once normalised) is just
\[
I(\delta) = 1 + \delta\log_2(\delta) + (1-\delta)\log_2(1-\delta)
\]

Information is transferred between phones by publishing a data payload along with a daily exposure key on a public server. The payload and exposure key is periodically downloaded by all mobiles, if any mobile has a PID in its log that can be derived from the exposure key then the payload is deemed delivered to that phone.

However, a daily exposure key will match all contacts in a given day, but in order to implement belief propagation we need to send different messages to different contacts on the same day. So somehow we need to deliver different messages to all contacts on a given day by publishing a single data payload. We want to do this in an efficient way that protects information about the contacts.

For forward messages we do this by publishing the phone's current belief in the posterior marginal infectiousness distribution for the day of the exposure key (under the approximation that the infectiousness of the person was constant over that period. In practice, we can take the infectiousness at the mean of the contact times for that day). From this single distribution all contacts can calculate their own forward messages.

To see this, suppose the posterior infectiousness distribution for the day in question is given by $\Phi_i(\iota)$. This is known to phone $i$. Consider now one of the same phone's Bernoulli clusters for this day $\Phi_{B}$, and suppose this is connected to cluster $\Phi_j$ on the other participant's phone. If there have been no backward messages from $\Phi_j$ to $\Phi_{B}$, the forward message from $\Phi_{B}$ to $\Phi_j$ will be equal to
\[
\delta = \int B(\tau_{a\rightarrow b}(t)|\iota)\Phi_i(\iota) d\iota = \int \iota\Phi_i(\iota) d\iota
\]
If on the other hand $\Phi_{B}$ has received a backwards message, $\delta'$,  from $\Phi_j$, we need to remove the effect of this backward message before calculating the forward message. The backward message to $\Phi_{B}$ would in turn create another backward message from $\Phi_{B}$ to $\Phi_i(\iota)$ with value
\[
\delta'\iota + (1-\delta')(1-\iota) = (1 - \delta') + (2\delta' - 1)\iota
\]
So, the posterior belief about $\iota$ must be
\[
\Phi_i(\iota) = \alpha((1 - \delta') + (2\delta' - 1)\iota)\Phi'_i(\iota)
\]
where $\Phi'_i(\iota)$ is the belief with the backward message removed and $\alpha$ is a normalising constant. So, the correct forward message from $\Phi_{B}$ to $\Phi_j$ must be
\begin{equation}
\delta = \int i \Phi'_i(\iota) d\iota = \frac{1}{\alpha}\int \frac{\iota\Phi_i(\iota)}{(1 - \delta') + (2\delta' - 1)\iota} d\iota
\label{forwardMessage}
\end{equation}
where
\begin{equation}
\alpha = \int \frac{\Phi_i(\iota)}{(1 - \delta') + (2\delta' - 1)\iota} d\iota
\label{normalisation}
\end{equation}
but since the receiving phone, $j$, knows the value of $\delta'$ (i.e. the backward message it sent) it can reconstruct the correct forward message from the Bernoulli cluster given only $\Phi_i(\iota)$.

So, if phone $i$ publishes a parameterised version of $\Phi_i(\iota)$ along with a daily exposure key, all contacts for that day can derive their individual forward messages. We suggest publishing $\Phi_i(\iota)$ in terms of its first $n$ moments. By taking the Taylor expansion of equations \ref{forwardMessage} and \ref{normalisation} about $\delta' = 0.5$, the forward message can be expressed as a series in the moments of $\Phi_i(\iota)$.

There is a similar problem with backward messages if we wish to publish them with the user's daily exposure key; different backward messages would need to be sent to contacts that happened on the same day.

However, suppose for all contacts in a day $\left<t_1,a_1,b_1,\iota_1\right>...\left<t_n,a_n,b_n,\iota_n\right>$, we imagine an additional cluster that sits between the the Bernoulli clusters on the contacts' phones, $ \Phi_B(\tau_{b_1\rightarrow a_1},\_) ... \Phi_B(\tau_{b_n\rightarrow a_n},\_)$, and the local $\Phi_i$ cluster. The new cluster aggregates all incoming transmission edges for one day into a single effective transmission using a logical OR operation, so that the effective transmission is true if any of the transmissions for that day are true. The $\Phi_i$ cluster now sends a single message to the OR cluster, $\delta_\Phi$, which is its posterior likelihood function of the aggregated transmission for the day, divided by any messages sent from the OR cluster. Since we're not interested in normalisation constants we'll assume that $\delta_\Phi$ has been normalised and so can be sent as a single number. Suppose also that that we've received forward messages $\delta_1...\delta_m$ from the remote Bernoulli clusters. The OR cluster needs to send the following backward message to Bernoulli cluster $j$
\[
\delta'_j = \frac{\delta_\phi}{\delta_\phi + (1-\delta_\phi)\prod_{i\ne j}(1-\delta_i) + \delta_\phi(1 - \prod_{i\ne j}(1-\delta_i))}
\]
\[
= \frac{\delta_\phi}{ (1-2\delta_\phi)\prod_{i\ne j}(1-\delta_i) + 2\delta_\phi}
\]
\[
= \frac{\delta_\phi}{ \frac{1-2\delta_\phi}{1-\delta_j}\prod_{i}(1-\delta_i) + 2\delta_\phi}
\]
where we've normalised the backward message for convenience even though it's a likelihood.

So, instead of sending separate backward messages to each Bernoulli cluster, we publish $\delta_\phi$ and $\prod_i (1-\delta_i)$ along with the daily exposure key. Then, since each phone already knows it's own $\delta_j$ (i.e. the forward message it passed for that contact) they can each calculate their own backward message from the published values. In this way we only need to make a single publication to send backward messages to all contacts on that day.

\subsection{Training the model: The parameter server}

``Epidemiological'' messages from a phone to the parameter server will consist of the likelihood function of the parameters for any new test results. If there are many parameters, an approximation could be sent in order to reduce bandwidth requirements; for example, by assuming independence between certain parameter sets. Since the parameter server doesn't need to know the identity of the sender in order to update its belief, the updates can be sent anonymously.

Forward messages from the parameter server to each phone could be sent by publishing the posterior belief in the parameters. Phones could update their values on a daily basis. In a similar way to that described above, if a user has sent a backward message to the parameter server, the effects of that message can be removed from the published posterior by simply dividing, in order to calculate the correct forward message from the parameter server.

The prior of the parameter server could also be updated centrally in response to new scientific understanding of the dynamics of the disease.

\section{Inference over multiple encounters between the same two people}

If every pair of people meet at most once, the belief propagation algorithm will result in beliefs that are equal to the Bayesian marginal posteriors. However, when the same two people have many encounters, this can lead to beliefs that do not correspond exactly to the Bayesian posterior. To demonstrate this, suppose A and B meet twice within a short space of time, and suppose A is infected. The variables of interest here are $\iota$, the infectiousness of A at the time of the meetings, $\epsilon(t)$, B's first exposure time and $\tau_1$ and $\tau_2$, the transmission events from A to B for each meeting. The joint of these is given by
\[
P(\iota, \tau_1, \tau_2, \epsilon(t)) = P(\iota)B(\tau_1|\iota)B(\tau_2|\iota)P(\epsilon(t)|\tau_1,\tau_2)
\]
where $B(\tau|\iota)$ is the Bernoulli distribution, and, from equation \ref{firstExposure}
\[
P(\epsilon(t)|\tau_1,\tau_2) = \delta(t-t_1)\tau_1 + \delta(t-t_2)\tau_2(1-\tau_1)
\]
where, $t_1$ and $t_2$ are the times of the encounters. In order to keep the demonstration simple, we set $\rho$ to zero and assume no other contacts, although the same argument applies in the presence of other contacts and non-zero $\rho$.

Starting with forward inference, the correct marginal for $P(\epsilon(t))$ is
\[
P(\epsilon(t)) = \int \delta(t-t_1)\iota + \delta(t-t_2)\iota(1-\iota) P(\iota) d\iota = \delta(t-t_1)\overline{\iota} + \delta(t-t_2)\left(\overline{\iota} - \overline{\iota^2}\right) 
\]
However, the belief propagation algorithm will end up with a belief
\[
B(\epsilon(t)) = \delta(t-t_1)\overline{\iota} + \delta(t-t_2)\overline{\iota}(1-\overline{\iota}) = \delta(t-t_1)\overline{\iota} + \delta(t-t_2)\left(\overline{\iota}-\overline{\iota}^2\right)
\]
So the the belief that the first encounter was a first exposure is correct, but the belief that the second encounter was a first exposure is in error by the variance of $P(\iota)$.

To give a sense of the size of error in reverse inference, we take the worst case scenario which is that B is certainly infected, and has only been in contact with $A$, so either $\tau_1$ or $\tau_2$ must be the first exposure. This is equivalent to saying $\tau_1 \vee \tau_2 = true$. The correct posterior marginal is
\[
P(\iota | \tau_1 \vee \tau_2 = true) = A\left(2\iota - \iota^2\right)P(\iota)
\]
where $A$ is a normalising constant. Expressing $\iota$ as a perturbation from the mean $\iota = \overline{\iota} + \delta$ and expanding in powers of $\delta$ gives
\[
P(\delta | \tau_1 \vee \tau_2 = true) = A\left(\overline{\iota}(2 - \overline{\iota}) + 2(1-\overline{\iota})\delta - \delta^2\right)P(\overline{\iota}+\delta)
\]
 However, the propagation algorithm would give, after one cycle of message passing, 
\[
B(\iota | \tau_1 \vee \tau_2 = true) = A' (\iota(1 - \overline\iota) + \overline{\iota})^2P(\iota)
\]
Expanding this in powers of $\delta$ gives
\[
B(\delta | \tau_1 \vee \tau_2 = true) = A'(\overline{\iota}(2 - \overline{\iota})) \left(
\overline{\iota}(2 - \overline{\iota}) +
2(1-\overline{\iota})\delta +
\frac{(1-\overline{\iota})^2}{\overline{\iota}(2 - \overline{\iota})}\delta^2
\right)P(\overline{\iota}+\delta)
\]
However, since $\overline{\iota}$ and $\delta$ will be small (of the order 0.01\cite{luo2020modes}), the expected error is in the order $\mathcal{O}\left(\frac{\overline{\iota^2}}{2\overline{\iota}}\right)$. Further loops of message passing would introduce higher order terms in $\delta$ but since $\delta$ is small these can be ignored.

We may wish to simply neglect these errors, or if we want more accuracy we could identify repeated encounters.

\subsection{Identifying repeated encounters}

If the phone receiving a message is able to identify transmission events that involve the same person, the corresponding Bernoulli clusters can be bundled into a single cluster and belief propagation will again be exact.

If the repeated encounters occur on the same day, the phone that receives the message can easily identify them since the daily exposure key will match multiple PIDs in its log. However, if the encounters were on different days, the matching is more difficult as they will have different daily exposure keys and, by design, these cannot be tracked to an individual.

Repeated encounters will most often be between members of the same household, close friends or work colleagues. Users could be given the option to match encounters with known contacts by allowing their phones to communicate directly in any convenient way (e.g. Bluetooth). Suppose person A has published a number of daily exposure keys. For any genuine close contacts with person B, both A and B know the daily exposure key and the PIDs that were exchanged during the encounter. If B's PIDs match multiple published exposure keys, B's phone can choose one, use the exchanged PIDs to encrypt the corresponding daily exposure key and send the encrypted message to A's phone. A's phone can then try to decrypt the message using the PID pairs corresponding to its encounters. If the decrypted message successfully matches A's daily exposure key, a message can be sent back to B confirming that the encounter was between A and B. In order for B's phone to keep the number of PID matches secret, it can send a number of dummy requests to make up the total number of requests to some randomly chosen number. In this way, the only information either phone can gain is which encounters involved the two parties. Since this is not a secret between the two parties in any case, no private information is revealed.

In order to match repeated encounters with strangers, the user could be prompted when the probability of two messages being from the same person goes above some threshold. The user could then use information about the times of the contacts, their memory of events and knowledge of their daily behaviour to try to work out whether they were repeated encounters. For example, if the user goes to the same coffee shop every morning, then the pattern of encounters at the same time every morning may allow them to assign them as repeated encounters.

\section{Deciding when to self-isolate, test and release}
Now we have a way of estimating posterior marginal probabilities, we can use these to make decisions. We adopt a decision theoretic approach by calculating the expected cost of each option and taking the one that has the lowest cost. We measure cost in days of life lost. If a person dies of COVID-19, the cost of that death is the expected number of extra days of life that person would have had, had they not contracted the disease.

From this, we can calculate the cost of someone becoming infected when $R<1$. In this case, a single infection will result in an average of $-\frac{1}{\ln(R)}$ total infections. If $P_d$ is the case fatality ratio, then we would expect $-\frac{P_d}{\ln(R)}$ deaths and an average of
\[
c_{id} = -\frac{P_dL}{\ln(R)}
\]
days of life lost, where $L$ is the average number of days lost when someone dies of COVID-19, considering age distribution and life expectancy.

There is also a cost, $\chi$, associated with a person being hospitalised for a day (this covers the personal, inter-personal and financial costs). If $\bar{h}$ is the average number of days of hospitalisation per infection then we have a total expected cost of a single infection
\[
c_{i} = -\frac{P_dL + \chi\bar{h}}{\ln(R)}
\]

Let the cost of a person having to self-isolate for a day be $c_s$ and let $c_{Tn}$ be the cost of taking the $n^{th}$ test. These are subjective values which must weigh up the social, personal and economic costs.

If we assume decisions are made on a daily basis then on each day we need to decide whether to self-isolate for that day and whether to take test $n \in 1...N$. Clearly, any tests with immediate results should be taken before the decision whether to self isolate for that day so the decision which tests to take should be made first. Given any test results, a person should self isolate on that day if
\[
c_s < \bar{\iota}(t)\rho_cc_i
\]
where $\rho_c$ is the user's average daily rate of close contacts and $\bar{\iota}(t)$ is the user's expected infectiousness on that day, given all the evidence to date.

If we decide to test, there is a fixed immediate cost of taking a test, $c_{Tn}$, but in return we gain information about the disease onset distribution. We can put a value on that information by calculating the difference in expected cost with and without that information. If having the information reduces our expected cost by more than the cost of obtaining the information then, on average, it is worth paying the price of the test for the information. So, the expected cost at time $t$ is
\[
\bar{c}(t) = \min(\bar{c_0}(t), c_{Tn} + P(\xi_n^+(t)|C(t))\bar{c}(\xi_n^+(t)) + P(\xi_n^-(t)|C(t))\bar{c}(\xi_n^-(t)))
\]
where $\bar{c}(\xi_n^+(t))$ and $\bar{c}(\xi_n^-(t))$ is the expected cost given a positive/negative test result and $\bar{c}_0(t)$ is the expected cost of not testing.
\[
\bar{c}_0(t) = (c_s - \bar{\iota}(t)\rho_cc_i)H(\bar{\iota}(t)\rho_cc_i - c_s) + \bar{c}(t+1)
\]
where $H(.)$ is the Heaviside step function.
\[
\bar{c}(\xi_n^-(t)) = (c_s - \bar{\iota}(t|\xi_n^-)\rho_cc_i)H(\bar{\iota}(t|\xi_n^-)\rho_cc_i - c_s) + \bar{c}(t+1)
\]
If we assume that information that raises the probability of a transmission event above some threshold, $p_t$ will trigger isolation and contact tracing of a close contact, then the expected cost of getting a positive test is
\[
\bar{c}(\xi_n^+(t)) = 
  (c_s - \bar{\iota}(t|\xi_n^+)\rho_cc_i)H(\bar{\iota}(t|\xi_n^+)\rho_cc_i - c_s) + 
  \sum_{A} c_a^\rightarrow(t-t_c) 
+   \sum_{B} c_a^\leftarrow(t-t_c) 
+ \bar{c}(t+1)
\]
where 
\[
A = \left\{
    \left<t_c,a,b\right> \in C(t): P(\tau_{a\rightarrow b}|\xi_n^+) > p_{t} \wedge P(\tau_{a\rightarrow b}) \le p_{t}
  \right\}
\]
\[
B = \left\{
    \left<t_c,a,b\right> \in C(t): P(\tau_{b\rightarrow a}|\xi_n^+) > p_{t} \wedge P(\tau_{b\rightarrow a}) \le p_{t}
  \right\}
\]
$\bar{c}_a^\leftarrow(\Delta t)$ is the expected cost saving of isolating and contact tracing an infector at a time $\Delta t$ after the contact and $\bar{c}_a^\rightarrow(\Delta t)$ is the expected cost saving of isolating and contact tracing an infectee at a time $\Delta t$ after the contact. These expected values can be pre-calculated using Monte-Carlo simulation.

Notice that calculation of $\bar{c}(t)$ requires calculation of $\bar{c}(t+1)$ so we have a recursion.

To make the recursion tractable, we make a few assumptions. First, since the logging of symptoms can be done with very little cost, we assume that symptoms should be logged every day unconditionally (in practice, this will consist of the logging of the onset of and recovery from any relevant symptoms, with the assumption of no symptoms otherwise). Secondly, we assume that we take at most one clinical test a day, and that the appropriate clinical test (i.e. PCR, antigen or antibody) can be decided by clinical considerations based on our belief about onset time and whether there have been previous positive tests (i.e. whether we are trying to show that a person has been infected or is safe to release).

With these assumptions, we can easily calculate the recursion 7 days into the future without putting a significant computational load on the phone. Since on each day either no test is performed, or a test is performed and returns positive, or a test is performed and returns negative, there are three scenarios to evaluate each day and $3^7 = 2187$ scenarios in total. Beyond the 7 day horizon, we assume that if isolated the user will remain isolated until no danger is posed or if not isolated, the user will remain out of isolation. Since we don't know what the result of symptom tests will be, we take a Monte-Carlo sample of symptom onset times (say 100 samples) and run each scenario with each sample, giving a total of just over $200,000$ scenarios to run. This is not a large computational load, but can be reduced further if desired by pruning highly unlikely scenarios (such as getting two consecutive negative tests at the peak of viral shedding after positive confirmation) and unreasonable testing schedules (e.g. performing a test around symptom onset if a positive test has already been obtained).

\section{Discussion and further work}

We have described a privacy-preserving way of doing distributed Bayesian inference on a network of mobile phones in order to inform decisions made by a contact tracing app. The algorithm also allows the disease model on each phone to be updated in response to the data collected during contact tracing and allows Epidemiologists to learn about the dynamics of the disease.

Further work is necessary to ensure that the network can withstand malicious attack, and a full analysis should be made to ensure that there is no way that the content of the messages could be used to identify a person. However, we have shown that Bayesian inference and epidemiological data collection is possible within the context of decentralised contact tracing as described in the Apple/Google specification.

\bibliographystyle{unsrturl}
\bibliography{inferenceRefs}

\end{document}